\documentstyle[epsfig, twocolumn]{iso98}

\newcommand{\thepapername}{}

\begin{document}

\setlength{\parindent}{0pt}
\setlength{\parskip}{ 10pt plus 1pt minus 1pt}
\setlength{\hoffset}{-1.5truecm}
\setlength{\textwidth}{ 17.1truecm }
\setlength{\columnsep}{1truecm }
\setlength{\columnseprule}{0pt}
\setlength{\headheight}{12pt}
\setlength{\headsep}{20pt}
\pagestyle{veniceheadings}

\title{\bf A DEEP AND ULTRA-DEEP ISOCAM COSMOLOGICAL SURVEY THROUGH \\
GRAVITATIONALLY LENSING CLUSTERS OF GALAXIES\thanks{ISO is an ESA
project with instruments funded by ESA Member States (especially the PI
countries: France, Germany, the Netherlands and the United Kingdom) and
with the participation of ISAS and NASA.}}

\author{{\bf L.~Metcalfe$^1$, B.~Altieri$^1$, B.~McBreen$^2$, J-P.~Kneib$^3$,
M.~Delaney$^2$, A.~Biviano$^4$, } \vspace{2mm} \\
{\bf M.F.~Kessler$^1$, K.~Leech$^1$, K.~Okumura$^5$, B.~Schulz$^1$, 
D.~Elbaz$^6$, H.~Aussel$^6$} \vspace{2mm} \\
$^1$ISO Data Centre, Astrophysics Division of ESA, Villafranca del Castillo,
P.O Box 50727, 28080 Madrid, Spain \\
$^2$Physics Department, University College Dublin, Stillorgan Road, Dublin 4,
Ireland \\
$^3$Observatoire Midi-Pyrenees, 14 Av. E. Belin, 31400 Toulouse, France \\
$^4$Osservatorio Astronomico di Trieste, via G.B. Tiepolo 11, 1-34131 Trieste, 
Italy \\
$^5$Institut d'Astrophysique Spatiale, Bat. 121, Universite Paris-Sud, F-91405
Orsay, France \\
$^6$DSM/DAPNIA/Sap, CEA-Saclay, 91191 Gif-sur Yvette Cedex, France.}

\maketitle

\begin{abstract}

We present imaging results and source counts from an ISOCAM deep and ultra-deep
cosmological survey through gravitationally lensing clusters of galaxies at 
7\,$\mu$m and 15\,$\mu$m. A total area of about 53\,sq.arcmin was covered in
maps of three clusters. The lensing increases the sensitivity of the survey. 
A large number of luminous mid-infrared (MIR) sources were detected behind the 
lenses, and most could be unambiguously identified with visible counterparts.
Thanks to the gravitational amplification, these results include the faintest 
MIR detections ever recorded, extending source counts to an unprecedented 
level. 
The source counts, corrected for cluster contamination and lensing distortion 
effects, show an excess by a factor of 10 with respect to the prediction of a 
no-evolution model, as we reported for A2390 alone in 
\cite*{a239099}. These results support the A2390 result that the 
resolved 7\,$\mu$m and 15\,$\mu$m background radiation intensities are
(1.7$\pm$0.5)$\times$10$^{-9}$ and (3.3$\pm$1.3)$\times$10$^{-9}$ 
W\,m$^{-2}$\,sr$^{-1}$ respectively, integrating from 30\,$\mu$Jy to 50\,mJy. 
\vspace {5pt} \\


  Key~words: ISO; infrared astronomy; gravitational lensing; galaxy clusters;
number counts; morphology.

\end{abstract}

\section{INTRODUCTION}
\label{sec:intro}

Following the identification of the first cosmological gravitational lens by
\cite*{young80} the theoretical and observational exploitation of the
lensing phenomenon developed rapidly. In 1987 luminous giant arcs were 
discovered in the fields of some galaxy clusters by \cite*{lp86}, and 
independently by \cite*{sou87}, and these were soon recognised to be Einstein 
rings (\cite{pac87}).
Since then observations of cluster-lenses have been extended to 
wavelengths other than the visible (e.g.\ \cite*{smail93} for the NIR, and 
\cite*{smail97} for the Sub-mm region).
At the same time models describing the lensing have been 
refined (\cite{jp96}) to the point where source counts made through a cluster 
lens can be corrected to yield the counts which would be found in the absence 
of the lens.

We report key results of a deep-survey programme, initiated almost a 
decade ago as part of the $^{\prime\prime}$Central Programme$^{\prime\prime}$  
of the ISO mission, and which has extended the exploitation of the 
gravitational lensing phenomenon to the Mid-infrared (MIR) for the first time 
by using ISOCAM (\cite{cc96}) on ESA's ISO spacecraft (\cite{MFK96}).

The programme achieved deep and ultra-deep (ISOCAM) imaging through a sample of
gravitationally lensing clusters of galaxies, benefiting from the lensing to 
search for background objects to a depth not otherwise achievable. 
Lensing amplifies sources and, for a given apparent flux limit, suppresses 
confusion, due to the apparent surface area dilation. To the flux-limits of this
sample, even rich foreground galaxy clusters are essentially transparent at 
15\,$\mu$m.

\section{OBSERVATIONS, DATA-REDUCTION AND DATA-ANALYSIS}
\label{sec:obsred}
\subsection{Observations}
\label{sec:obs}

We observed the fields of the well-known gravitationally lensing
galaxy clusters Abell 370, Abell 2218 and Abell 2390. A total area of about
53 sq.arcminutes was covered. Table~\ref{tab:table} lists the observational 
parameters of the survey which used CAM's 6.7\,$\mu$m (LW2) and 14.3\,$\mu$m
(LW3) filters. By employing the CAM 3\,arcsec per-pixel field-of-view (PFOV)
and raster step 
sizes which were multiples of 1/3 pixel, over many raster steps, we 
achieved a final mosaic pixel size of 1\,arcsec.  The diameter of the PSF 
central maximum is, in
arcseconds, 0.84$\times\lambda(\mu m)$, and the FWHM is about half that. 
The final 1\,arcsec pixel size improved source separation and allowed a better 
cross-identification with observations at other wavelengths.
This strategy differs from many other CAM deep surveys which used the 6\,arcsec
per-pixel field-of-view. The 5$\sigma$ sensitivity 
limits are given in Table~\ref{tab:table} and depend on the lensing 
cluster in question. These limits refer to apparent source brightness {\bf 
before} correction for the effects of lensing. For the clusters observed, the 
highest-magnification regions of the lenses give amplifications of 5 to 10 - so
with this method we could in principle make 5$\sigma$ detections in LW2 and 
LW3, of sources with intrinsic brigtnesses as low as 6$\mu$Jy and 13$\mu$Jy 
respectively, if any such faint background source falls in the regions of 
strongest lensing. 
Typical magnifications are about 2 over the central few square-arcmin of the 
clusters.

Lensing also causes a surface dilation effect over the area probed. This 
spatial dilation is stronger towards the core of the cluster and increases with 
source-plane redshift. These effects need to be corrected during analysis in
order to compare results with {\it blank} sky counts
(e.g.\ in the Hubble Deep field and Lockman Hole -- \cite{row97}; \cite{tan97};
\cite{herv98}; \cite{elb98}).

\begin{table*}[!ht]
 \caption{\em Observational parameters and lensing clusters of galaxies used 
 for the survey. On-chip integration time (t$_{int}$) was always 5.04 seconds 
 and the 3\,arcsec per-pixel field-of-view was used.
}
  \label{tab:table}
   \begin{center}
    \leavevmode
     \footnotesize
      \begin{tabular}{ccccccccccc}
       \hline
        \noalign{\smallskip}
 FIELD      & FILTER\,$^{\dagger}$ & Readouts & Raster            &  
Raster            &   dm   &   dn   &   area         &  depth             &  
No. of times&  Tot. time  \\
            &                      & per step & steps X           &    
 steps Y          & arcsec & arcsec & (sq.arcmin)    & 5$\sigma\,(\mu$Jy) &  
repeated    &   (sec.)    \\
        \noalign{\smallskip}
        \noalign{\smallskip}
 Abell 2390 & LW2                  &   13     &   10              &  
 10               &   7    &  7     &    5.3         &   33              &  
  4         &  29300      \\
        \noalign{\smallskip}
            & LW3                  &   13     &   10              &  
 10               &   7    &  7     &    5.3         &   67              &   
  4         &  29300      \\
        \noalign{\smallskip}
        \noalign{\smallskip}
 Abell 2218 & LW2                  &   14     &   12              &  
 12               &  16    & 16     &   16           &      $*$          &  
  2         &  22000      \\
        \noalign{\smallskip}
            & LW3                  &   14     &   12              &  
 12               &  16    & 16     &   16           &  150              &  
  2         &  22000      \\
        \noalign{\smallskip}
        \noalign{\smallskip}
 Abell  370 & LW2                  &   10     &   14              &  
 14               &  22    & 22     &   31.3         &      $*$          &  
  2         &  22688      \\
        \noalign{\smallskip}
            & LW3                  &   10     &   14              &   
 14               &  22    & 22     &   31.3         &  330              &   
  2         &  22688      \\
        \noalign{\smallskip}
        \hline
      \end{tabular}
   \end{center}
(*) in \cite*{met99}.
($\dagger$) LW2 and LW3 filters have reference wavelengths 6.7 and 
14.3\,$\mu$m, respectively. 
\end{table*}

\vskip -10pt

\subsection{Data reduction and analysis}
\label{sec:analyse}

\vskip -2pt
The reduction of ISO faint source data is extremely challenging and many 
work-years have been invested in developing effective and reliable techniques. 
The ultimate performance is particularly sensitive to the effectiveness with 
which the detector's global responsive transient and cosmic-ray-induced 
glitches can be removed from
the data. Two substantially independent data reduction approaches were applied.
These are described in detail in the ISOCAM Faint Source Report 
(\cite{faint98} - available on the ISO www page). We particularly refer to 
Sections 3.3 and 3.4 of that document, and related 
sections. Key techniques employed to perform photometry and to extract 
source-lists and sensitivity limits are described in Section 5 
of the report and also in \cite*{a239099}.

Detailed lensing models of the three clusters used have been produced by 
\cite*{jp96} and (1999), and \cite*{bez99}.
By correcting for the lens magnification and surface dilation effects, 
contamination by cluster galaxies (mainly at 7\,$\mu$m), and non-uniform 
sensitivity of our maps, we can derive number counts to compare with blank sky 
counts. The object density per flux bin was computed using 
magnification-dependant surface areas derived from the lensing models so
dividing the maps into sets of sub-maps. The unreliable data at the boundaries 
of the maps were discarded.

\section{RESULTS}
\label{sec:results}

\vskip -2pt
In Section~\ref{sec:obs} the potential limiting sensitivity of this 
survey was described (6 and 13\,$\mu$Jy at 5$\sigma$ in LW2 and LW3
respectively). Achieving this potential is a matter of luck with respect to
source distribution behind the lens. In practice, the faintest 15\,$\mu$m
source detected was intrinsically an 18\,$\mu$Jy source amplified to 
80\,$\mu$Jy (a 6$\sigma$ detection).

Fake-source simulations are described in \cite*{a239099}, and show that
our 80\% completeness levels in LW3 (for which we report source counts here) 
before accounting for lensing are : 500, 250 and 100$\mu$Jy\ for A370, A2218 
and A2390, respectively. Our simulations for A370 and A2218 are still 
preliminary so their associated completeness levels may be adjusted slightly 
when the full analysis is finished. Lensing has a variable effect over the 
field so it is not possible to state a global completeness level after 
correction for lensing. Rather, we stopped counting at the 50\% completeness 
flux level in the highest magnification bins we could use. This was an 
intrinsic 33\,$\mu$Jy at 15\,$\mu$m for A2390, corresponding to a 5$\sigma$ 
threshold for counting a source.

\vskip -2pt
\subsection{Cluster images at 15\,$\mu$m}
\label{sec:images}

\begin{figure}[!ht]
   \begin{center}
   \leavevmode
   \centerline{\epsfig{file=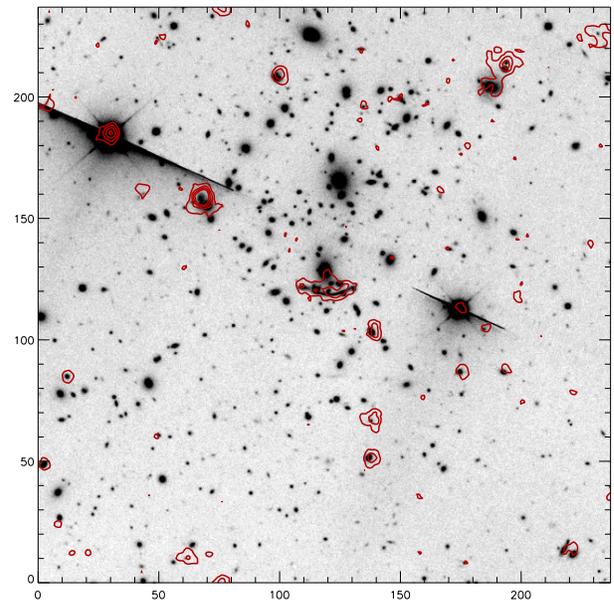,
               width=8.0cm}}
   \end{center}
 \caption{\em Contours of an ISOCAM 15\,$\mu$m image of Abell 370 over a deep 
CFHT I-band image. The giant arc is well detected by ISO, along with a number 
of other MIR sources in the field, some with quite faint optical counterparts.
(All axes are arcseconds.)}
 \label{fig:a370_lw3}
 \end{figure}

\begin{figure}[!ht]
   \begin{center}
   \leavevmode
   \centerline{\epsfig{file=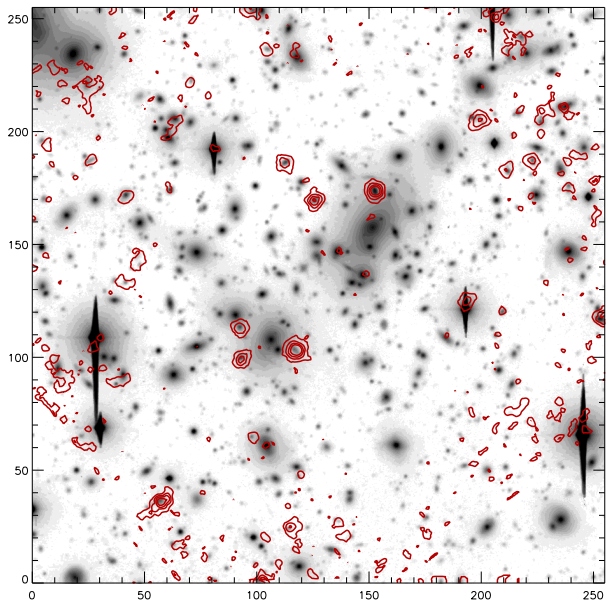,
               width=8.0cm}}
   \end{center}
 \caption{\em Contours of the inner 4\,$\times$\,4 square arcmin of an ISOCAM 
15\,$\mu$m image of Abell 2218 over a deep Palomar 5m I-band image (Smail, 
private communication). Note the strong 
ISO detection of the giant arc at (117,105), the several MIR sources 
corresponding to faint optical counterparts [e.g.\,(200,205), (45,140), 
(105,240)] and the fact that the IR 
$^{\prime\prime}$arclets$^{\prime\prime}$ are generally not from the same 
population as the known visual arclets.}
 \label{fig:a2218_lw3}
 \end{figure}

\begin{figure}[!ht]
   \begin{center}
   \leavevmode
   \centerline{\epsfig{file=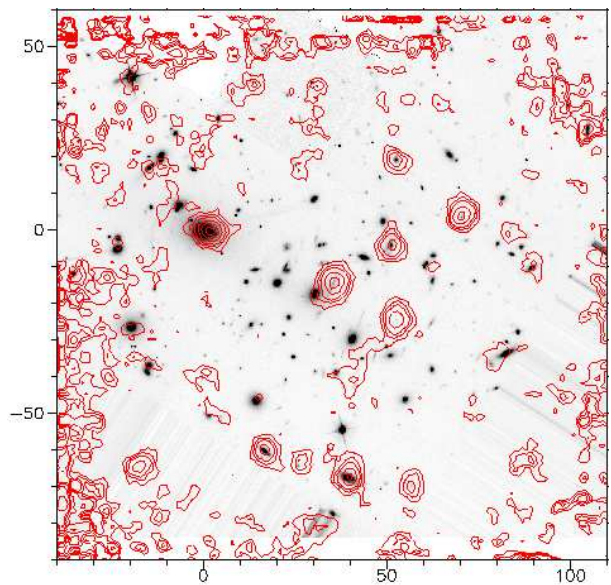,
               width=8.0cm}}
   \end{center}
 \caption{\em Contours of an ISOCAM 15\,$\mu$m image of Abell 2390 over an HST 
(F814W) I-band image. Note the ISO 
detection of part of the straight arc near the centre of the image (35,-15), 
and the two strong ISO detections (spectroscopic redshift unknown) at
(53,-25) and (70,05). These objects have extremely red colours and very 
faint visual fluxes. Their colours, if interpreted as due to redshift, imply 
photometric redshifts greater than 3.
}
 \label{fig:a2390_lw3}
 \end{figure}

\vskip -2pt
Overlays onto visible images of ISOCAM LW3 (15 $\mu$m) deep images through the
three clusters can be found 
in figures\,~\ref{fig:a370_lw3},\,~\ref{fig:a2218_lw3} 
and\,~\ref{fig:a2390_lw3}.
(See also \cite*{barv99} for additional ISO data on A2218.)

From these images we can remark that we generally do not see the optical 
arc(lets) at 15\,$\mu$m. We see a Large Number of MIR sources, shown from
spectroscopic, photometric or lensing redshifts to be behind the lens. These
are, in effect, IR arc(let)s.

Thanks to our high-resolution images we have been able to unambiguously 
identify 90\% of sources with (sometimes extremely faint) counterparts 
in deep NIR and visual (HST/WFPC2 and ground-based) images. Spatial resolution 
is good enough to allow recognition of IR correlations with 
the visual morphology of objects. (E.g.\,The ISO detection peak matches the 
bright visual knot in the straight arc (\cite{Pel91}) near the centre of the 
A2390 map.)

We found that at 7\,$\mu$m about half the sources are cluster galaxies, the 
other half are lensed background sources. We infer that almost all 15\,$\mu$m 
sources are lensed background galaxies because most sources for which we have 
spectroscopic redshifts are, while the photometric and lensing-inversion 
redshifts for the remaining sources are consistent with 
locations beyond the cluster.  This indicates that, at 15\,$\mu$m, the 
cluster-core becomes transparent, similar to the Sub-mm case (\cite{smail97}). 

Some ISO targets have extremely red optical and NIR colours.  It appears that 
15\,$\mu$m imaging favours
selection of star-forming galaxies and dusty AGNs that are not evident in 
UV/optical surveys.

 \begin{figure}[!ht]
   \begin{center}
   \leavevmode
   \centerline{\epsfig{file=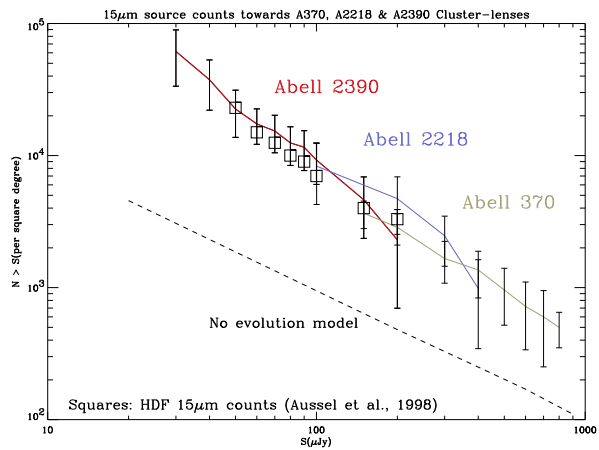,
               width=8.0cm}}
   \end{center}
 \caption{\em ISO 15\,$\mu$m source counts for three clusters. The slope from
A2390 is -1.5\,$\pm$\,0.3 and N$_{15}$($>$30\,$\mu$Jy) = 
13\,$\pm$\,5\,arcmin$^{-2}$ at 15\,$\mu$m. Counts for A2390 are fully 
corrected for lensing. Counts for A370 and A2218 have only been roughly 
corrected. Planned final corrections are not expected to significantly alter 
the slope, but could move A370 and A2218 data-points slightly along the curve 
towards lower flux.}
 \label{fig:counts}
 \end{figure}

\subsection{Number counts and source properties}
\label{sec:counts}

We have extracted 15\,$\mu$m source counts for the three clusters taking only
5$\sigma$ detections into account for statistical studies. The analysis 
performed for A370 and A2218 is provisional, and is being refined. It is 
expected, however, that the final results will differ only in detail. We find 
a total of 71 sources within our survey area (see Table~\ref{tab:table}), with
30, 23 and 18 from A2390, A2218 and A370 respectively.

The 15\,$\mu$m counts show a steadily increasing excess by a factor of 
order 10 over no-evolution models (see Figure~\ref{fig:counts}, and
\cite{fran94}).
These results confirm the Lockman hole result of 
\cite*{elb98}, and agree with the HDF counts of \cite*{herv98}. There is no 
sign of flattening of the counts at the faintest levels. The slope stays close 
to -1.5\,$\pm$\,0.3 down to the 30\,$\mu$Jy limit. These source densities 
favour extreme evolution models.

Integrating the number counts over our A2390 flux range (30\,$\mu$Jy to 
200\,$\mu$Jy)
and over other ISOCAM surveys (\cite{elb98}) up to 50\,mJy we find that
the resolved background light is $(3.3 \pm 1.3) \times 10^{-9}$ and
$(1.7\pm 0.5)\times 10^{-9}$ W\,m$^{-2}$\,sr$^{-1}$ at 15\,$\mu$m and 
7\,$\mu$m, respectively.
The 15$\mu$m lower limit is close to the upper limit from the
gamma-CMBR photon-photon pair production (\cite{sf97}).

\section{CONCLUSIONS}
\label{sec:conclude}

The method of looking through gravitational lenses served to extend ISO 
coverage of background sources to the deepest levels, especially at 15\,$\mu$m. 
The results suggest that abundant star formation occurs in very dusty 
environments at z$\sim$1. 
Caution must be employed to take account of the effects of dust in inferring 
global star formation history from UV or optical luminosities of high-z 
galaxies. 
We will present a more complete description of these MIR observations,
including 7\,$\mu$m and 15\,$\mu$m photometry. in a forthcoming paper 
(\cite{met99}).
A detailed study of the optical properties of the MIR galaxies may reveal the
nature of the evolving or new population of luminous IR sources.

\section*{ACKNOWLEDGMENTS}

{\scriptsize
The ISOCAM data presented in this paper was analysed using ``CIA",
a joint development by the ESA Astrophysics Division and the ISOCAM
Consortium. The ISOCAM Consortium is led by the ISOCAM PI, C. Cesarsky,
Direction des Sciences de la Matiere, C.E.A., France.
}


\begin{thebibliography}{}

\bibitem[\protect\astroncite{Altieri et~al.}{1998}]{faint98}
Altieri,~B., Metcalfe,~L., Ott,~S. et al.\  {\hskip 8pt} 1998, 
\verb!http://www.iso.vilspa.esa.es/users/
expl_lib/CAM_list.html!

\bibitem[\protect\astroncite{Altieri et~al.}{1999}]{a239099}
Altieri,~B., Metcalfe,~L., Kneib,~J-P. et al.\  1999, A\&A, in press, 
astro-ph/9810480

\bibitem[\protect\astroncite{Aussel et~al.}{1998}]{herv98}
Aussel, H., Cesarsky, C.J., Elbaz, D., Starck, J.-L.\ 1998, A\&A in press,
astro-ph/9810044

\bibitem[\protect\astroncite{Barvainis et~al.}{1999}]{barv99}
Barvainis., R., Antonucci R., Helou., G.\ 1999, in preparation.

\bibitem[\protect\astroncite{Bezecourt et~al.}{1999}]{bez99}
Bezecourt., J., Kneib, J.-P., Soucail, G., Ebbels, T.\ 1999, submitted to 
A\&A, astro-ph/9810199

\bibitem[\protect\astroncite{Cesarsky et~al.}{1996}]{cc96}
Cesarsky, C.J., Abergel, A., Agnese, P. et al.\ 1996, A\&A 315, L32

\bibitem[\protect\astroncite{Elbaz et~al.}{1998}]{elb98}
Elbaz, D., Aussel, H., Cesarsky, C. J. et al.,\ 1998, Proc. of 34th Liege
Int. Astr. Coll. on the NGST, astro-ph/9807209

\bibitem[\protect\astroncite{Franceschini et~al.}{1994}]{fran94}
Franceschini, A., Mazzei, P., De Zotti, G., Danese, L.\ 1994, ApJ 427, 140

\bibitem[\protect\astroncite{Kessler et~al.}{1996}]{MFK96}
Kessler, M.F., Steinz, J.A., Anderegg, M. et al.\ 1996, A\&A 315, L27
 
\bibitem[\protect\astroncite{Kneib et al.}{1996}]{jp96}
Kneib, J-P., Ellis, R.S., Smail, I., Couch, W.J., Sharples, R.M.\ 1996, ApJ 471,
643

\bibitem[\protect\astroncite{Kneib et al.}{1999}]{jp99}
Kneib, J-P., et al.\ 1999, in preparation

\bibitem[\protect\astroncite{Lynds and Petrosian}{1986}]{lp86}
Lynds, R., Petrosian, V.\  1986, Bull. Am. Astr. Soc. 18, 1014

\bibitem[\protect\astroncite{Metcalfe et~al.}{1999}]{met99}
Metcalfe, L., et al.\ 1999, in preparation

\bibitem[\protect\astroncite{Paczyn\'ski}{1987}]{pac87}
Paczyn\'ski, B.\  1987, Nat 325, 572

\bibitem[\protect\astroncite{Pell\'o et al.}{1991}]{Pel91}
Pell\'o, R., LeBorgne, J.F., Soucail, G., Mellier, Y., Sanahuja, B., 1991, ApJ
366, 405

\bibitem[\protect\astroncite{Rowan-Robinson et~al.}{1997}]{row97}
Rowan-Robinson, M., Mann, R., Oliver, S. et al.\ 1997, MNRAS 289, 490

\bibitem[\protect\astroncite{Smail et~al.}{1993}]{smail93}
Smail, I., Ellis, R., Arago\'n-Salamanca, A. et al.\  1993, MNRAS 263, 628

\bibitem[\protect\astroncite{Smail et~al.}{1997}]{smail97}
Smail,~I., Ivison,~R.J., Blain,~A.W.\  1997, ApJ 490, L5


\bibitem[\protect\astroncite{Soucail et~al.}{1987}]{sou87}
Soucail, G., Mellier, Y., Fort, B., Mathez, G., Cailloux, M.\  1987, 
The Messenger, 50, 5

\bibitem[\protect\astroncite{Stanev and Franceschini}{1997}]{sf97}
Stanev, T., Franceschini, A.\ 1997, ApJ 494, L159

\bibitem[\protect\astroncite{Taniguchi et~al.}{1997}]{tan97}
Taniguchi, Y., Cowie, L., Sato, Y. et al.\ 1997, A\&A, 328, L9

\bibitem[\protect\astroncite{Young et~al.}{1980}]{young80}
Young, P., Gunn, J.E., Oke, J.B., Westphal, J., Kristian, J.\  1980, ApJ 241, 
507


\end{thebibliography}
\end{document}